\documentclass[structabstract]{aa}  

\usepackage{enumerate}
\usepackage{graphicx}
\usepackage{txfonts}
\usepackage{natbib}

\newcommand{\boldnabla}{\mbox{\boldmath$\nabla$}}

\begin{document}
  
  \title{Dynamical Relaxation of Coronal Magnetic Fields}

  \subtitle{III. 3D Spiral Nulls}

  \author{Jorge Fuentes-Fern\'andez and Clare E. Parnell}

  \institute{School of Mathematics and Statistics, University of St Andrews, North Haugh, St Andrews, Fife, KY16 9SS, Scotland}

  \date{}

  \abstract
  {The majority of studies on stressed 3D magnetic null points consider magnetic reconnection driven by an external perturbation, but the formation of a genuine current sheet equilibrium remains poorly understood. This problem has been considered more extensively in two-dimensions, but lacks a generalization into 3D fields.}
  {3D magnetic nulls are more complex than 2D nulls and the field can take a greater range of magnetic geometries local to the null. Here, we focus on one type and consider the dynamical non-resistive relaxation of 3D spiral nulls with initial spine-aligned current. We aim to provide a valid magnetohydrostatic equilibrium, and describe the electric current accumulations in various cases, involving a finite plasma pressure.}
  {A full MHD code is used, with the resistivity set to zero so that reconnection is not allowed, to run a series of experiments in which a perturbed spiral 3D null point is allowed to relax towards an equilibrium, via real, viscous damping forces. Changes to the initial plasma pressure and other magnetic parameters are investigated systematically.}
  {For the axi-symmetric case, the evolution of the field and the plasma is such that it concentrates the current density in two cone-shaped regions along the spine, thus concentrating the twist of the magnetic field around the spine, leaving a radial configuration in the fan plane. The plasma pressure redistributes in order to maintain the current density accumulations. However, it is found that changes in the initial plasma pressure do not modify the final state significantly. In the cases where the initial magnetic field is not axi-symmetric, a infinite-time singularity of current perpendicular to the fan is found at the location of the null.}
  {}

  \maketitle


\section{Introduction}

Three-dimensional magnetic null points have been studied in detail within the last decade in the main context of three dimensional magnetic reconnection. Their importance in solar and magnetospheric environments have been observationally identified by many authors, for example, in solar flares \citep{Fletcher01,Masson09}, in solar active regions \citep{Ugarte07} or at the Earth's magnetotail \citep{Xiao06}. However, a complete understanding of the formation of current layers in three-dimensional magnetic null points, through a physical dynamical relaxation, is still to be achieved, either mathematically or phenomenologically.

In two dimensions, current sheet formation and current accumulations have been widely studied around X-points both analytically \citep[e.g.][]{Dungey53,Green65,Somov76,Vekstein93,Craig94,Bungey95} and numerically, including the effects of plasma pressure \citep[e.g.][]{Rastatter94,Craig05,Pontin05b,Fuentes11}. For a more comprehensive review of current sheet formation in two dimensions see \citep{Priest00} and \citep{Biskamp00}.

All these studies assume the magnetic field is line -tied at the boundaries. This assumption imposes the constraints that energy and flux cannot leave or enter the system, and is reasonable since throughout the solar corona, most of the magnetic field lines are anchored in the photosphere. Furthermore, line-tying, along with ideal relaxation ensures that helicity must be conserved during the relaxation \citep{Moffat85} and mass is conserved within flux tubes.

In three dimensions, the processes of current accumulation and reconnection in three dimensions are significantly different to, and much more complex than, those in two-dimensions at X-type null points \citep[e.g.][]{Hesse88,Priest03}. In general, in three-dimensions, currents can accumulate and magnetic reconnection can occur either at nulls or in the absence of them. Away from magnetic null points, magnetic reconnection may take place in different structures, such as separators \citep{Longcope96,Longcope01,Haynes07,Parnell08,Parnell10a,Parnell10b} and quasi-separatrix layers \citep{Priest95,Demoulin96,Demoulin97,Aulanier06,Restante09,Wilmot09}. On the other hand, locally at null points in three dimensions, magnetic reconnection can occur in several different regimes \citep{Pontin04,Pontin05a,Masson09,Alhachami10,Pontin11,Priest09,Masson12}.

The nature of the reconnection that can take place around a three-dimensional null depends directly on the flows and the boundary disturbances \citep{Rickard96,Priest09}. One particular example of these reconnection regimes is {\it fan-spine reconnection}, where a shearing of the spine or the fan, drives the collapse of the null point. That is, the resulting Lorentz forces act in the same direction as the initial disturbance, thus increasing it and resulting in a folding of the spine and fan towards each other. The resulting reconnection takes place in the vicinity of the null, and implies that magnetic flux is transferred across the different topological regions of the system \citep[see][]{Pontin05a,Pontin07a,Pontin07b,Pontin07c}.

Another example of 3D null reconnection, and the one relevant for this paper, is {\it torsional reconnection}, where the flows are such that the spine and fan plane remain perpendicular to each other, and the field lines are twisted around the spine in the same direction above and below the fan ({\it torsional spine reconnection}) or in opposite directions above and below the fan ({\it torsional fan reconnection}). The resulting reconnection of field lines takes place about the spine or the fan, respectively, causing a slippage of the magnetic field through the plasma in a direction opposite to that of the twist, dissipating the current density. These regimes do not involve flow across the spine or the fan, and hence, the global topology of the field remains unchanged \citep[see][]{Bulanov02,Pontin04,Pontin07c,Wyper10,Pontin11}.

In most cases that focus specifically on the field about the null, they also assume that the initial field is symmetric about the axis of the spine. However, such an assumption does not need to be made \citep{Parnell96}. \citet{Alhachami10} and \citet{Pontin11} relax this assumption showing that the degree of asymmetry changes the rate of reconnection, but not its nature.

\citet{Parnell97} proved that the magnetic field locally about a non-potential 3D null point, i.e. the linear field about a non-potential null, produces a Lorentz force that cannot be balanced by a plasma pressure force. So there are no static equilibrium models of linear non-potential or force-free nulls. This means that in 3D, the relaxation towards an equilibrium magnetic field very close to a magnetic null point has the same choices that a 2D linear null has, i.e., i) to evolve towards a potential null or ii) to develop a current singularity (current accumulation) at the null. \citet{Klapper97} proves analytically that in 2D, in the absence of plasma, the collapse of an X-type null results in the current building up at the null forming an infinite-time singularity. Numerical experiments of null collapse in non-zero beta plasmas show that an infinite-time singularity is still found \citep{Craig05,Pontin05b,Fuentes11}.

In general terms, most of the studies mentioned already consider magnetic reconnection at an initially potential 3D null driven by some external force. Reconnection, however, is likely to occur at the location where current has accumulated. For instance, in the case of spontaneous reconnection initiated by microinstabilities, it can occur in an equilibrium current density structure. The aim of this paper is to investigate the nature of current accumulations at a 3D magnetic null, after a torsional-type disturbance. The current structures will be formed via the ideal magnetohydrodynamic (MHD) evolution of the field and the plasma. The nature of non-force-free equilibrium found after a non-resistive MHD evolution of a stressed 3D null point, and the effects of the plasma pressure in the formation of current layers, to our knowledge, have not been widely studied. \citet{Pontin05b} analysed the collapse of 3D nulls after a shearing-type perturbation (fan-spine collapse). In particular, they studied the formation of a current singularity at the location of the null in a non-force-free equilibrium, in an equivalent manner to the two-dimensional singularities studied by \citet{Craig05}. However, the formation of a static non-force-free equilibrium after a torsional-type disturbance has not yet been studied. We carry out such a study here.

Here, we consider the ideal evolution of a spiral null (non-potential) with an initial homogeneous current density parallel to the spine line, which may be thought of as the result of a torsional-spine disturbance. In particular, we are interested in the current accumulations that arise when a non-force-free equilibrium is reached. We evaluate the effects of the plasma pressure in the evolution, as both the initial disturbance and the background plasma pressure are changed systematically. This work is a continuation of the work carried out in \citet{Fuentes10,Fuentes11}, on non-resistive MHD relaxation of magnetic fields embedded in non-zero beta plasmas, using viscous forces to drive the relaxation, and allowing heat to be transferred to the plasma via the physical viscous heating term. Departures from axial symmetries are studied, obtaining current accumulations of a different nature, such as the formation of an infinite time singularity in some cases (which does not exist in the axi-symmetric case).

The paper is structured as follows. In Sec. \ref{sec2}, we present the equations that define the initial three-dimensional configuration and in Sec. \ref{sec3} we present details of the numerical experiments. In Sec. \ref{sec4} we present the results for 3D axi-symmetric nulls with initial spine-aligned current, and, in Sec. \ref{sec5} we present the results for non-symmetric nulls, and evaluate the formation of a current singularity at the location of the null. Finally, we conclude with a general overview of the problem in Sec. \ref{sec6}.


\section{Magnetic field configurations} \label{sec2}

\begin{figure}[b]
  \centering
  \vspace{0.3cm}
  \includegraphics[scale=0.80]{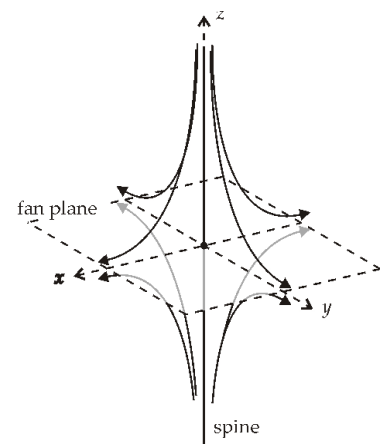}
  \caption{General topological structure of a 3D null point, including the spine line (thick) and fan plane (dashed). Field lines (thin) on either side of the fan plane run along the spine towards the null and then spread out just above or below the fan plane.}
  \label{fig:3dnull}
\end{figure}

The basic structure of a three-dimensional null point is illustrated in Fig. \ref{fig:3dnull}. The topological structure involves the spine line and the fan plane. In a positive 3D null (such as the one in the figure), the field lines approach the null parallel to the spine, and recede parallel to the fan plane. A negative null would show the opposite behaviour.

Following the description given in \citet{Parnell96} for linear 3D nulls, the magnetic field, ${\bf B}$, close to a null point may be expressed as
\begin{equation}
{\bf B}={\bf M}\cdot{\bf r}\;, \label{BMr}
\end{equation}
where ${\bf M}$ is a matrix with elements $M_{i,j}=\partial B_i/\partial x_j$, and ${\bf r}$ is the position vector $(x,y,z)^T$, which is likely to be small for general nulls where linearity may be weak. For linear nulls the current density is constant, and it can be divided into two components, parallel and perpendicular to the spine line. In the particular cases studied in this paper, the only non-zero component of the initial current density will be the one parallel to the spine, so that
 \begin{equation}
{\bf j}=\frac{1}{\mu}(0, 0, j_{sp})\;.
\end{equation}
Where, we have chosen a coordinate system in which the spine lies along the $z$-axis \citep{Parnell96}. In this case, the matrix ${\bf M}$ can be reduced to
\begin{equation}
{\bf M}=
\left( \begin{array}{ccc}
\frac{\partial B_x}{\partial x} & \frac{\partial B_x}{\partial y} & \frac{\partial B_x}{\partial z} \\
\frac{\partial B_y}{\partial x} & \frac{\partial B_y}{\partial y} & \frac{\partial B_y}{\partial z} \\
\frac{\partial B_z}{\partial x} & \frac{\partial B_z}{\partial y} & \frac{\partial B_z}{\partial z} \end{array} \right)\sim
\left( \begin{array}{ccc}
1 & -\frac{1}{2}j_{sp} & 0 \\
\frac{1}{2}j_{sp} & b & 0 \\
0 & 0 & -(b+1)\end{array} \right). \label{M_general}
\end{equation}
where $0<b<\infty$, in order to ensure that the spine of the null is along the $z$-axis, and that the null is positive.

For such non-potential nulls, the associated Lorentz force has no $z$-component and simply acts in planes parallel to the fan plane of the null. Furthermore, ${\bf j}\times{\bf B}={\bf 0}$ on $x=y=0$, hence, in our final non-force-free equilibrium, there will be no flux transferred across the fan, and this will remain perpendicular to the spine, but the geometry of the field lines and the distribution of current density and plasma will change.


\section{Numerical scheme} \label{sec3}

For the numerical experiments studied in this paper, we have used Lare3D, a staggered Lagrangian-remap code, which is second order accurate in space and time,  with user controlled viscosity, that solves the full MHD equations, with the resistivity set to zero \citep[see][]{Arber01}. The staggered grid is used to build conservation laws and maintains $\boldnabla\cdot{\bf B}=0$ to machine precision, by using the Evans and Hawley's constrained transport method for the magnetic flux \citep{Evans88}. The numerical domain is a 3D box with a uniform grid of $512^3$ points.

Magnetic field lines are line-tied at the boundaries and all components of the velocity are set to zero on the boundaries. The other quantities have their derivatives perpendicular to each of the boundaries set to zero. Hence, the quantities that are conserved over the whole domain are total energy and total mass. Since the field is frozen to the plasma (there is no diffusion to within numerical limits), the mass in a single flux tube (or along a field line) must be conserved.

The numerical code uses the normalised MHD equations, where the normalised magnetic field, density and lengths,
\begin{eqnarray*}
x=L\hat{x}\;,\;\;\;y=L\hat{y}\;,\;\;\;{\bf B}=B_n\hat{\bf B}\;,\;\;\;\rho=\rho_n\hat{\rho}\;,
\end{eqnarray*}
imply that the normalising constants for pressure, internal energy, current density and plasma velocity are,
\begin{eqnarray*}
p_n=\frac{B_n^2}{\mu}\;,\;\;\;\epsilon_n=\frac{B_n^2}{\mu\rho_n}\;,\;\;\;j_n=\frac{B_n}{\mu L}\;\;\;{\rm and}\;\;\;{\rm v}_n=\frac{B_n}{\sqrt{\mu \rho_n}}\;.
\end{eqnarray*}
The subscripts $n$ indicate the normalising constants, and the {\it hat} quantities are the dimensionless variables used in the code. The expression for the plasma beta can be obtained from this normalization as
\begin{eqnarray*}
\beta=\frac{2\hat{p}}{\hat{B}^2}\;.
\end{eqnarray*}
In this paper, we will work with the normalised quantities, but the {\it hat} is removed from the equations for simplicity.

The (normalised) equations governing our ideal MHD dynamical processes are
\begin{eqnarray}
\frac{\partial \rho}{\partial t}+\boldnabla\cdot(\rho{\bf v}) &=& 0\;,\label{n_mass}\\
\rho\frac{\partial{\bf v}}{\partial t}+\rho({\bf v}\cdot\boldnabla){\bf v} &=& -\boldnabla p + (\boldnabla\times{\bf B})\times{\bf B} + {\bf F}_{\nu}\;,\label{n_motion}\\
\frac{\partial p}{\partial t}+{\bf v}\cdot\boldnabla p &=& -\gamma p \boldnabla\cdot{\bf v}+H_{\nu}\;,\label{n_energy}\\
\frac{\partial{\bf B}}{\partial t} &=& \boldnabla\times({\bf v}\times{\bf B})\;,\label{n_induction}
\end{eqnarray}
where ${\bf F}_{\nu}$ and $H_{\nu}$ are the terms for the viscous force and viscous heating. The internal energy, $\epsilon$, is given by the ideal gas law, $p=\rho\epsilon(\gamma-1)$, with $\gamma=5/3$.

Finally, the time-scale in our experiments may be defined as the fast mode crossing time, $t_F$, i.e. the time for a fast magnetosonic wave to travel from the null to one of the boundaries (for instance in the $y$-direction), which is calculated as
\begin{equation}
t_F=\int_{y=0}^{y=1}\!\frac{{\rm d}y}{c_F(y)}\;,
\end{equation}
where $c_F(y)=\sqrt{v_A^2+c_s^2}$ is the local fast magnetosonic speed.


\section{Axi-symmetric nulls} \label{sec4}


\subsection{Initial state}

We first look at the relaxation of initial configurations of magnetic null points with constant current density everywhere in the direction parallel to the spine, of the form $(0,0,j_{sp})$, and we assume axial symmetry in the initial field by making $b=1$ in Eq. (\ref{M_general}). The magnetic field is then given by
\begin{equation}
(B_x, B_y, B_z)=(x-\frac{j_{sp}}{2}y, \frac{j_{sp}}{2}x+y\;, -2z)\;.
\end{equation}
The fan is perpendicular to the spine and lies in the $z=0$ plane. We have run various experiments with different initial pressure. Figure \ref{fig:sp_initial} shows the magnetic configuration of the initial state, for $j_{sp}=1$ and $p_0=1$. The magnetic field lines show a homogeneous twist about the spine, and the field lines lying in the fan define a logarithmic spiral. The only initially non-zero force in the fan plane is the magnetic tension force, which during the relaxation, will act to straighten the magnetic field lines. This force is such that, in principle, the system can evolve towards a potential configuration (a radial null) if energy and helicity is allowed to leave the system.


\subsection{Final equilibrium state}

\begin{figure*}[t]
  \begin{minipage}[b]{1.0\linewidth}

    \begin{minipage}[b]{0.50\linewidth}
      \centering
      \includegraphics[scale=0.32]{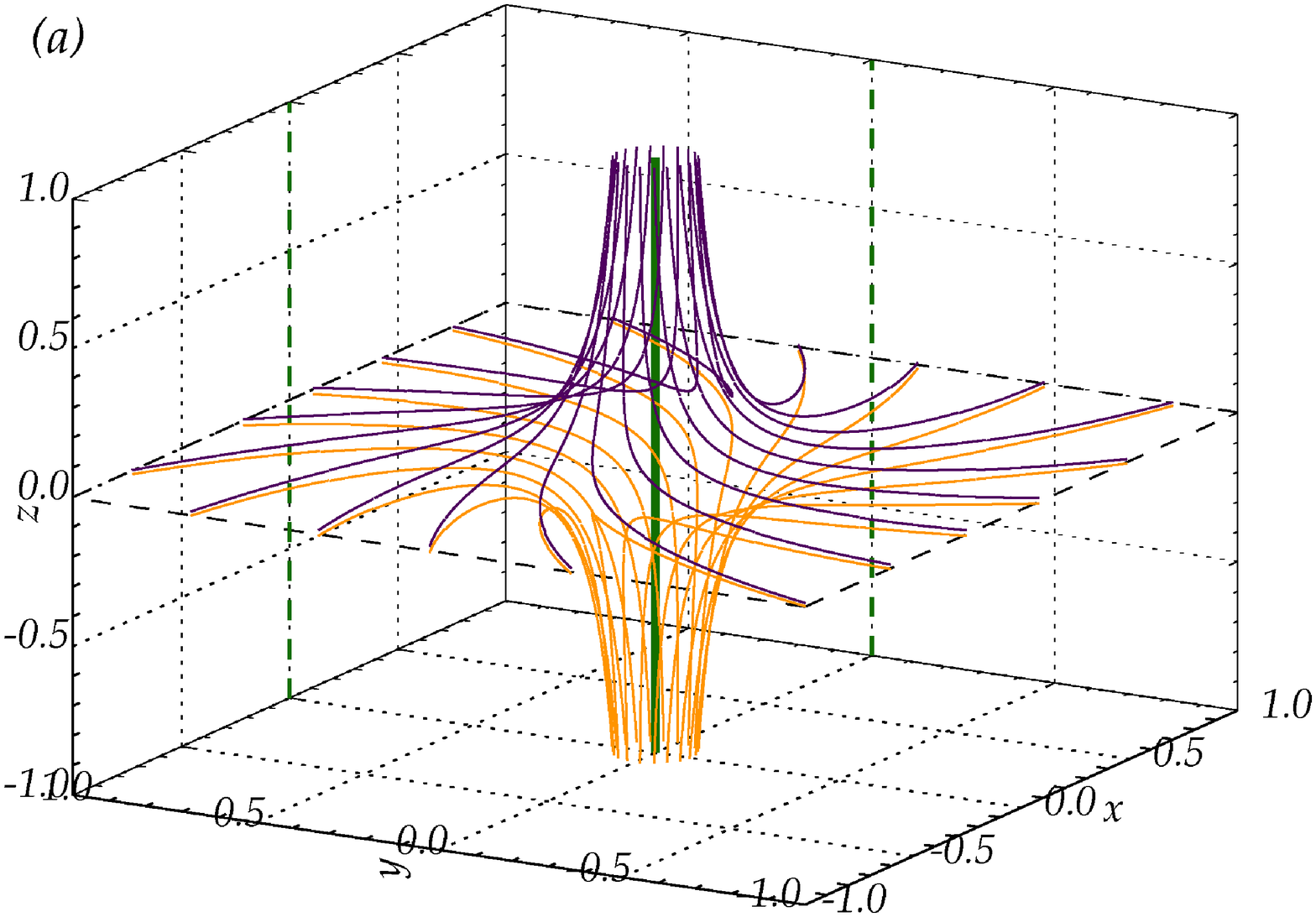}
    \end{minipage}
    \begin{minipage}[b]{0.50\linewidth}
      \centering
      \includegraphics[scale=0.30]{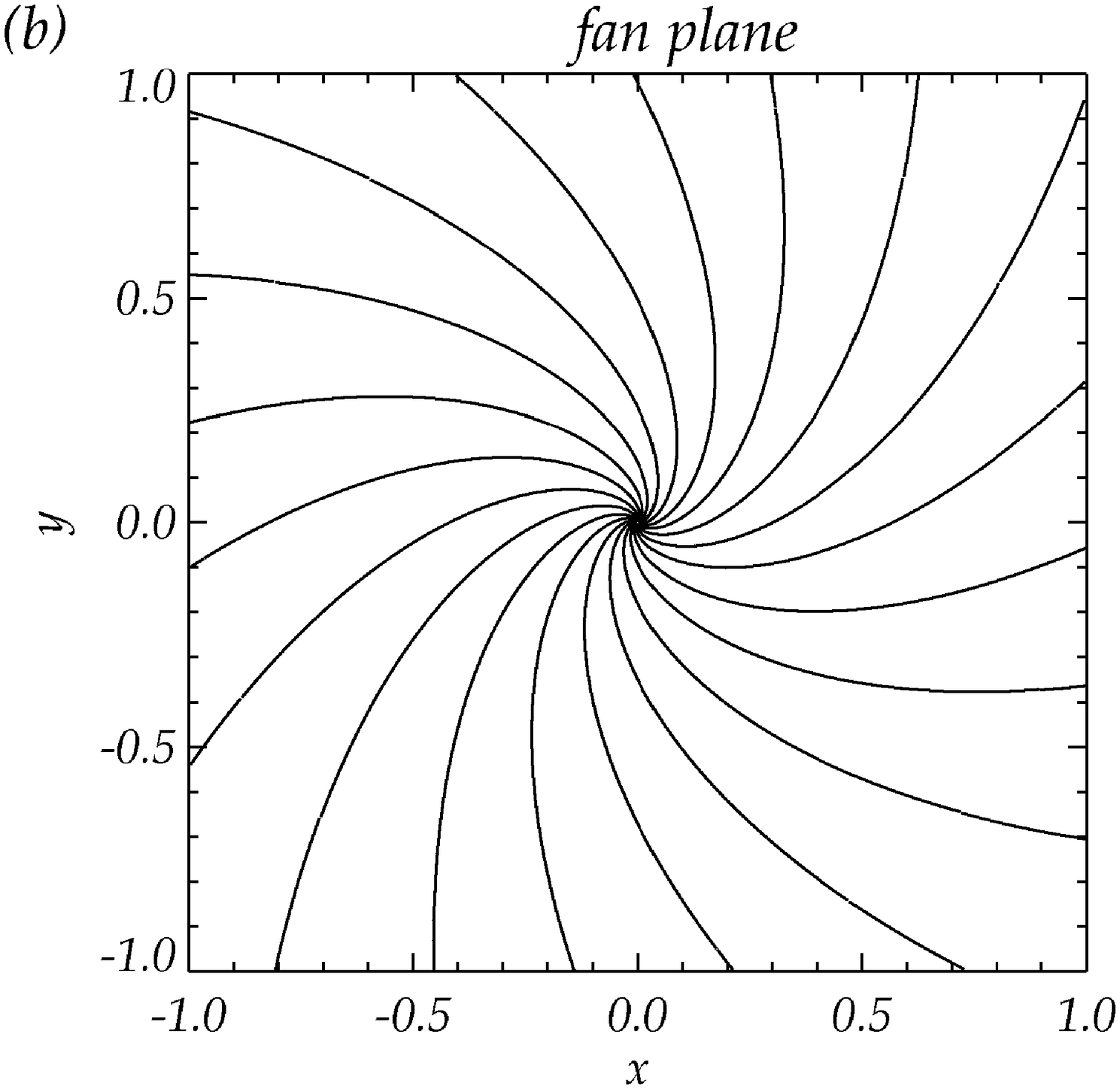}
    \end{minipage}

    \caption{Magnetic configuration for the initial non-equilibrium state with homogeneous spine-aligned current, for the case with $b=1$, $j_{sp}=1$ and $p_0=1$, showing (a) the 3D configuration with field lines above and below the fan in purple and orange respectively. The fan plane is outlined in dashed black and the spine is represented in green, with projections onto the $xz$-plane and $yz$-plane (dashed green lines). In (b), the field lines in the fan plane are plotted.}
    \label{fig:sp_initial}
  \end{minipage}
  \vspace{0.3cm}

  \begin{minipage}[b]{1.0\linewidth}

    \begin{minipage}[b]{0.50\linewidth}
      \centering
      \includegraphics[scale=0.32]{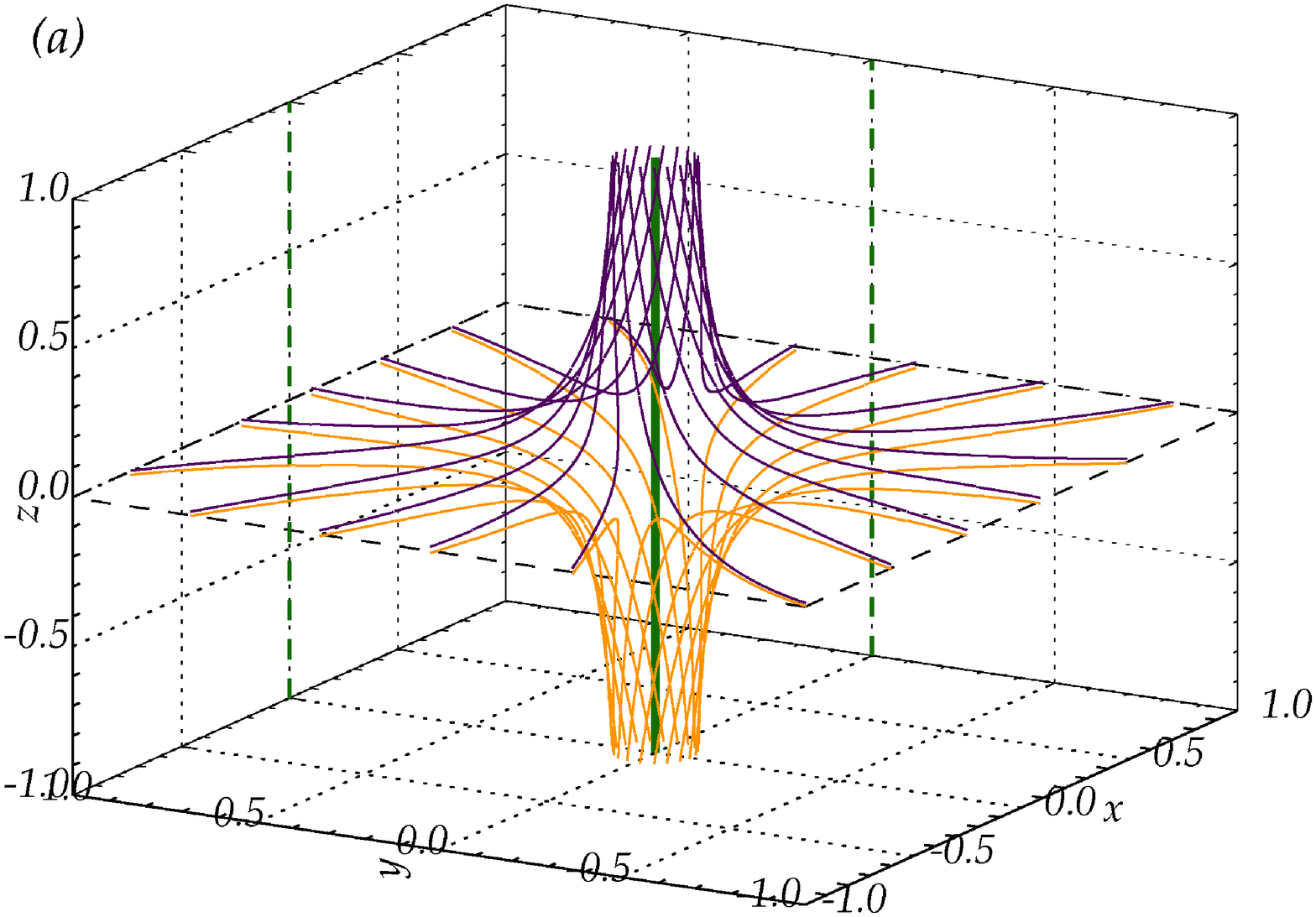}
    \end{minipage}
    \begin{minipage}[b]{0.50\linewidth}
      \centering
      \includegraphics[scale=0.30]{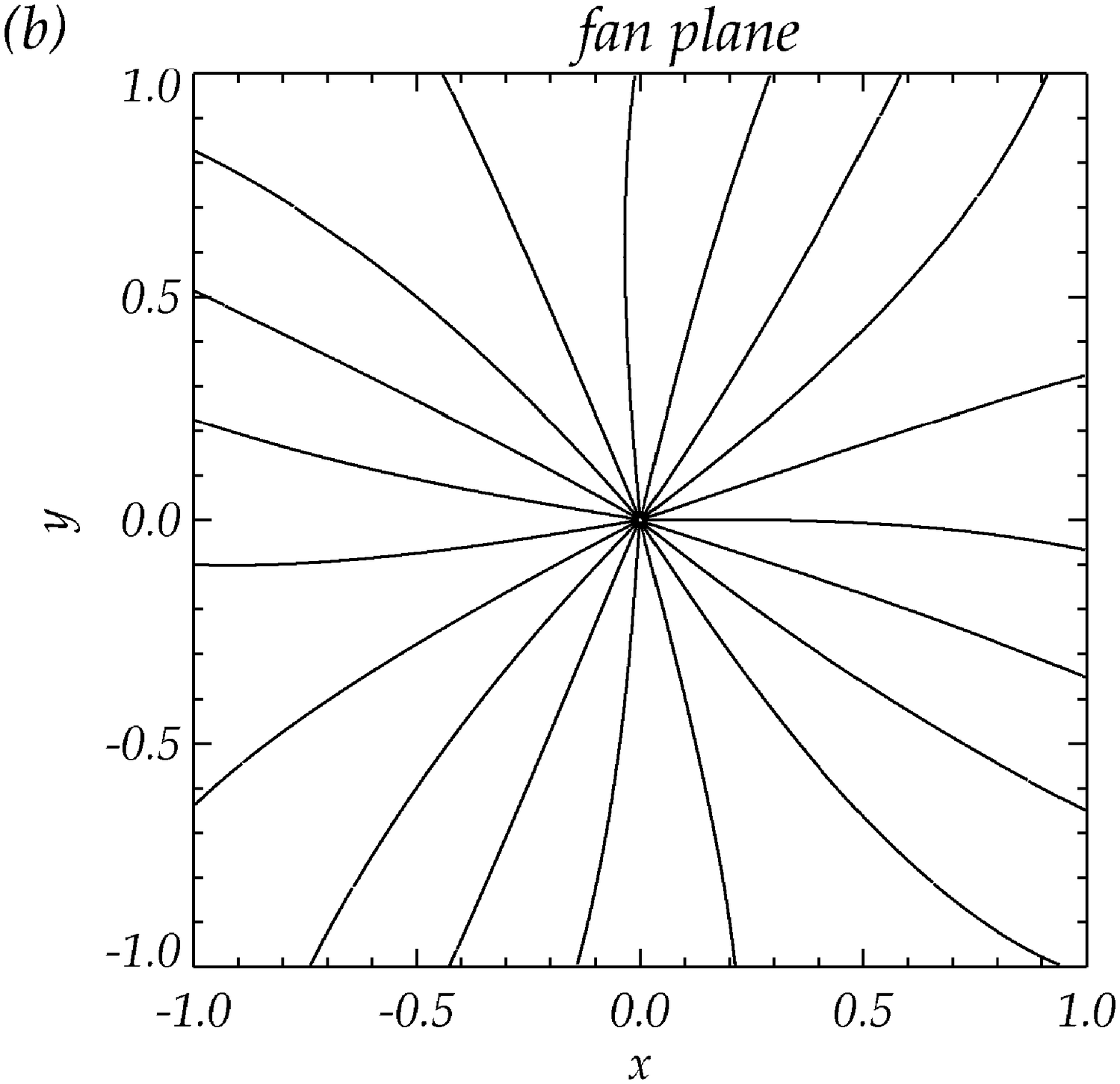}
    \end{minipage}

    \caption{Magnetic configuration for the final equilibrium state for the spiral null, for the same experiment as in Fig. \ref{fig:sp_initial}, showing (a) the 3D configuration with field lines above and below the fan in purple and orange, respectively. The fan plane is outlined in dashed black and the spine is represented in green, with projections onto the $xz$-plane and $yz$-plane in dashed green lines. In (b), the field lines in the fan plane are plotted.}
    \label{fig:sp_final}
  \end{minipage}
  \vspace{0.3cm}
\end{figure*}

We concentrate on the case shown in Fig. \ref{fig:sp_initial}, which has $j_{sp}=1$ and $p_0=1$. The final state is achieved after more than 300 fast magnetosonic crossing times. The magnetic field configuration at the end of the simulation is shown in Fig. \ref{fig:sp_final}. Owing to the line-tied boundaries, and the restriction of an ideal evolution, the field cannot dissipate the original twist. Instead, the relaxation appears to concentrate it onto the portions of field lines that are about the spine, above and below the null.

As can be seen in Fig. \ref{fig:sp_final}b, the field lines expand radially from the null and then bend slightly as they reach the boundaries of the domain. This bending is a consequence of the fact that the magnetic field lines are line-tied in a way that is not axi-symmetric. In Sec. \ref{sec:2halfD}, we consider a 2.5D approximation assuming cylindrical symmetry in a region close to the null point.

The total energy of the evolution is checked and is found to be conserved throughout the relaxation to within an error of 0.03\%. This conservation of energy demonstrates that undesirable effects, such as numerical diffusion, do not play a significant role in the relaxation. By the end of the experiment, only 1\% of the initial magnetic energy has been transferred into internal energy of the plasma, due to viscous damping. At this point, the velocities are effectively zero in the whole domain.

\begin{figure*}[t]
  \begin{minipage}[b]{1.0\linewidth}
    \centering
    \includegraphics[scale=0.60]{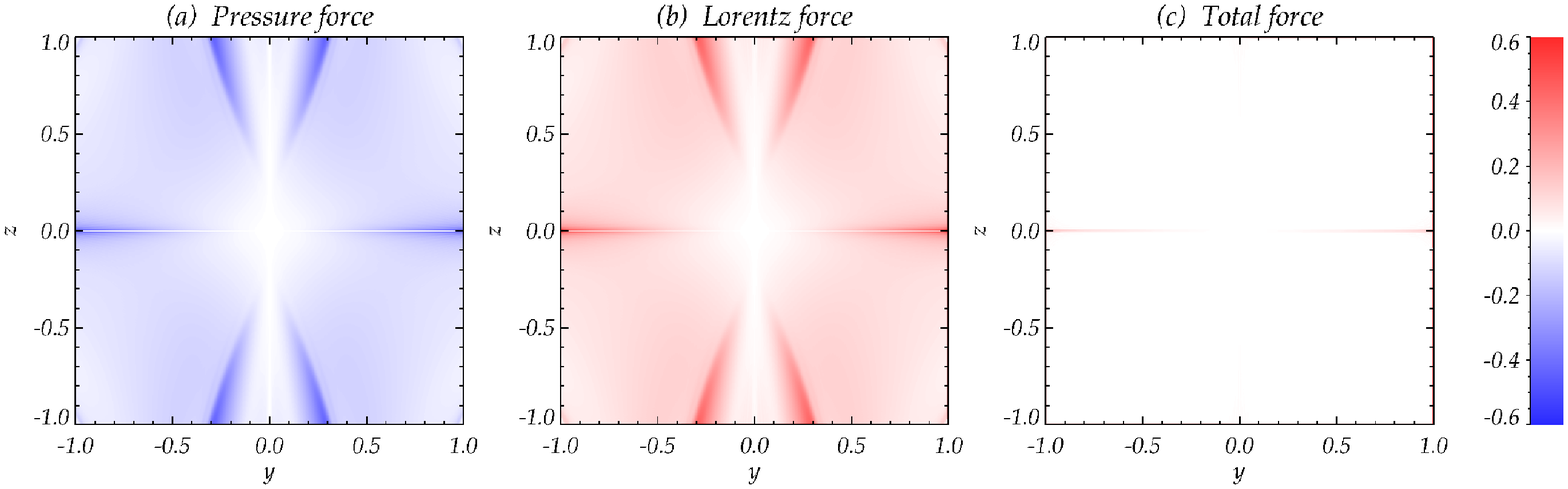}
    \caption{Contour plots of the different forces acting in the final equilibrium state in the $x=0$ plane, for the same experiment as in Fig. \ref{fig:sp_initial}. Showing, from left to right, the magnitude of the pressure force ($-|\boldnabla p|$), of the Lorentz force and of the total force. Values are normalised to the maximum force of the initial state. It can be observed that the pressure and Lorentz forces balance each other creating a non-force-free equilibrium.}
    \label{fig:sp_forces}
  \end{minipage}
  \vspace{0.3cm}

  \begin{minipage}[b]{1.0\linewidth}

    \begin{minipage}[b]{0.49\linewidth}
      \centering
      \includegraphics[scale=0.33]{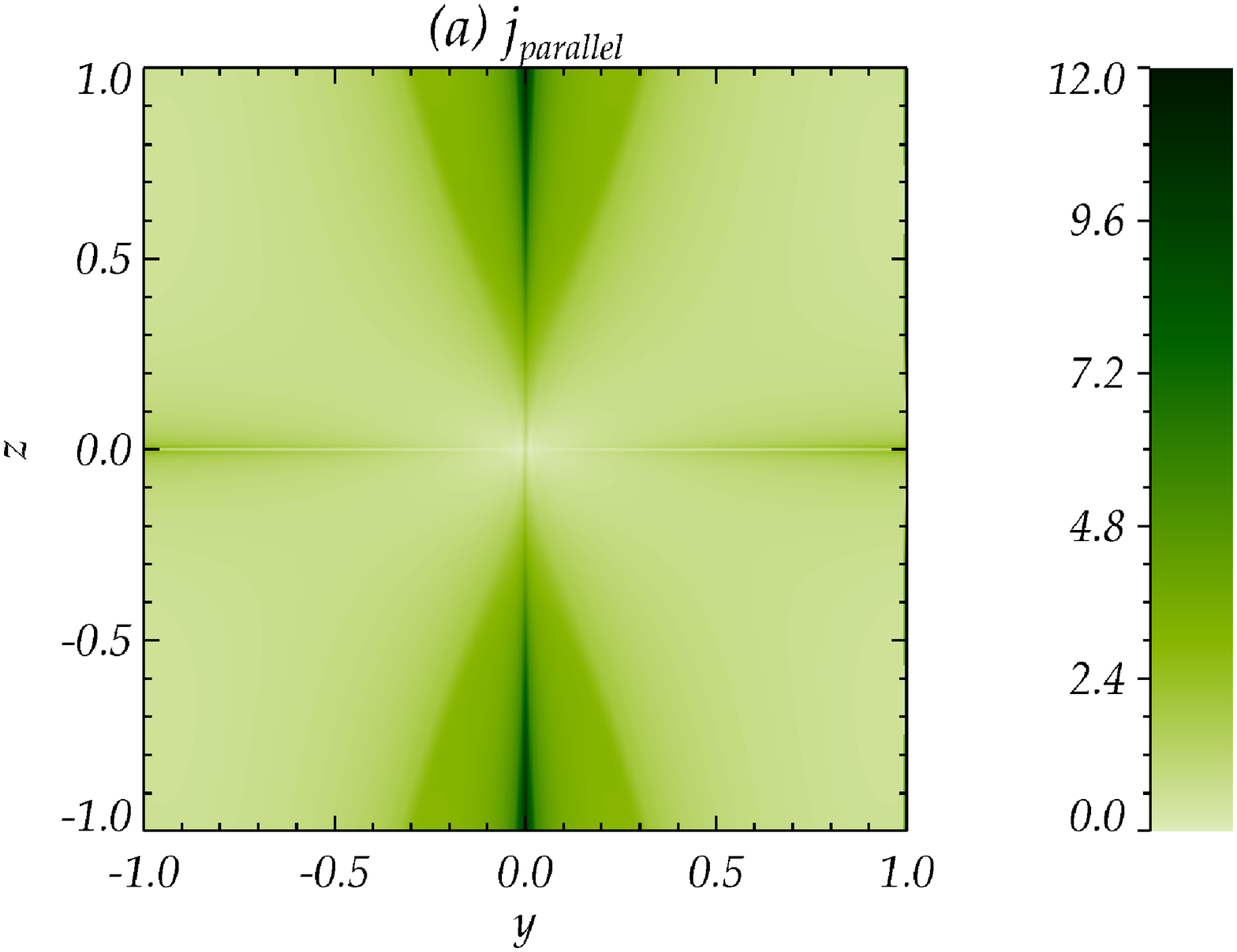}
    \end{minipage}
    \hspace{0.02\linewidth}
    \begin{minipage}[b]{0.49\linewidth}
      \includegraphics[scale=0.33]{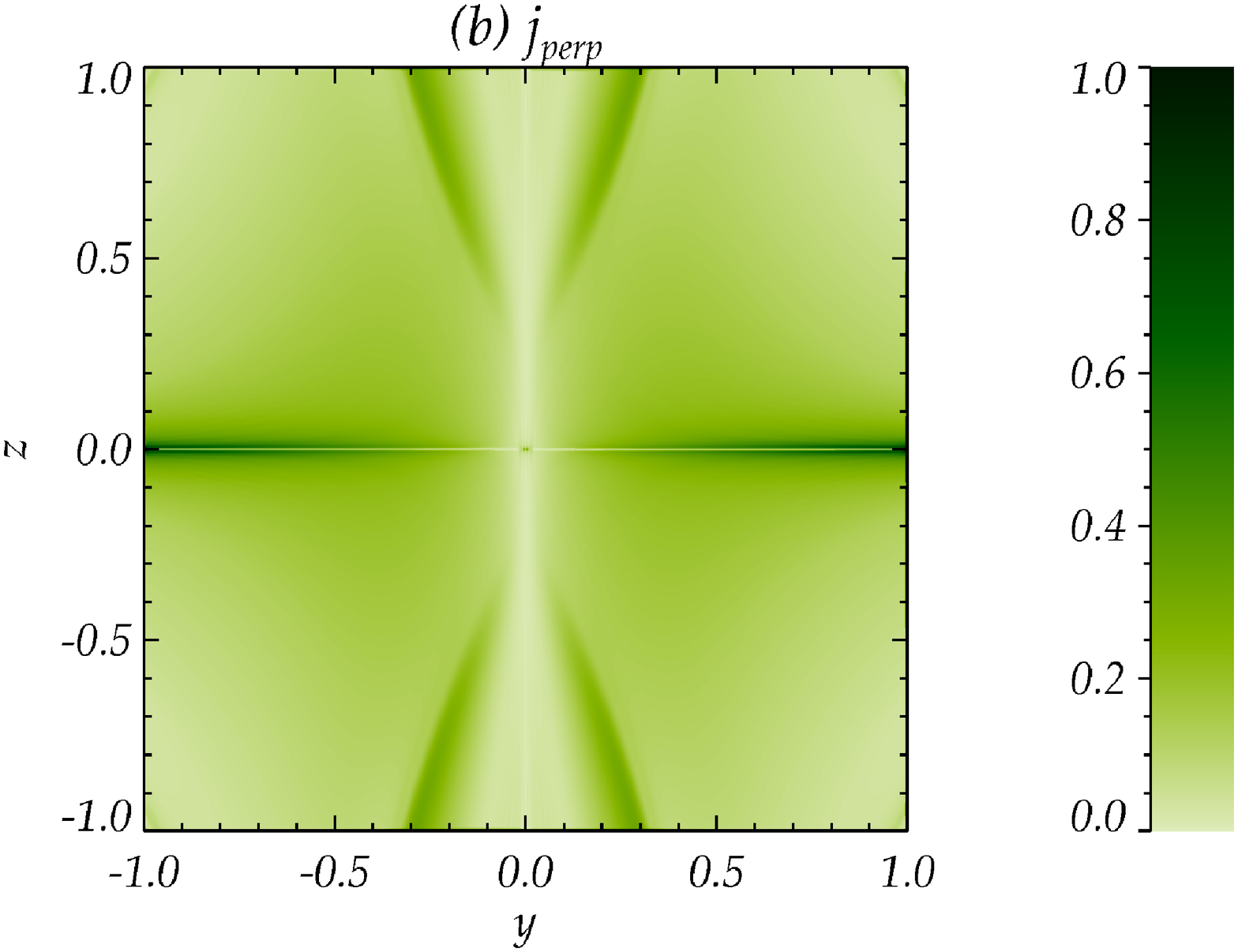}
    \end{minipage}
    
    \caption{Contour plots of the final equilibrium current density vector for a vertical cross section in the $x=0$ plane, for the same experiment as in Fig. \ref{fig:sp_initial}. The current has been split in two components: (a) parallel and (b) perpendicular to the magnetic field.}
    \label{fig:sp_current}
    
  \end{minipage}
  \vspace{0.3cm}
\end{figure*}

In order to establish the nature of the final equilibrium, we evaluate the forces at the end of the relaxation. Fig. \ref{fig:sp_forces} shows the different forces, namely, the Lorentz force, the pressure force and the total force (the sum of the two) in the final equilibrium state. The system is only force-free locally close to the null (where it is effectively potential) and along the spine line. Everywhere else, non-zero Lorentz forces are balanced by non-zero pressure gradients in a non-force-free equilibrium. This is the result of both the closed boundary conditions, which does not allow magnetic flux to be transferred through the boundaries, and the full MHD relaxation, which allows the different forces to balance each other. Note, the Lorentz force is such that, if the field lines were free to move at the boundaries, they would untwist, expelling their free magnetic energy and helicity out of the system, hence, achieving a potential state.

Around the axis of the spine, away from the null, a non-force-free envelope has formed (Fig. \ref{fig:sp_forces}a and b). These regions get larger and further apart as we move away from the null. Also, regions of {\it non-force-free-ness} appear about the plane of the fan, away from the null. Only small residual forces remain on the fan close to the boundaries (Fig. \ref{fig:sp_forces}c).

A non-force-free equilibrium implies an inhomogeneous distribution of the current and the plasma pressure. The final distributions of electric current density, parallel and perpendicular to the magnetic field, are shown in Fig. \ref{fig:sp_current}. These two components account for the force-free and the non-force-free contributions to the final equilibrium, respectively. The parallel component (Fig. \ref{fig:sp_current}a) is one order of magnitude larger than the perpendicular component (Fig. \ref{fig:sp_current}b), indicating that the equilibrium is not far from a force-free state.

At the exact location of the null point, the current density remains with its initial value of $j_{sp}=1.$ This is not inconsistent with \citet{Parnell97}, as the current is zero everywhere else. From the beginning of the experiment, at the location of the null itself, the Lorentz forces are zero, and so are the pressure forces.

The system is potential (i.e. current free) in a very localised region about the null point, where $r<0.1$, as expected from \citet{Parnell97}. The parallel current is mainly concentrated along the spine line, forming an hourglass shape centered on the null and lying along this axis (Fig. \ref{fig:sp_current}a). As already mention the field is force-free in these regions. The perpendicular current is enhanced at the edges of the hourglass cones, and about the fan plane (Fig. \ref{fig:sp_current}b). Naturally, these are the locations of the non-zero Lorentz and pressure forces (Fig. \ref{fig:sp_forces}a and b).

\begin{figure}[t]
  \centering
  \includegraphics[scale=0.61]{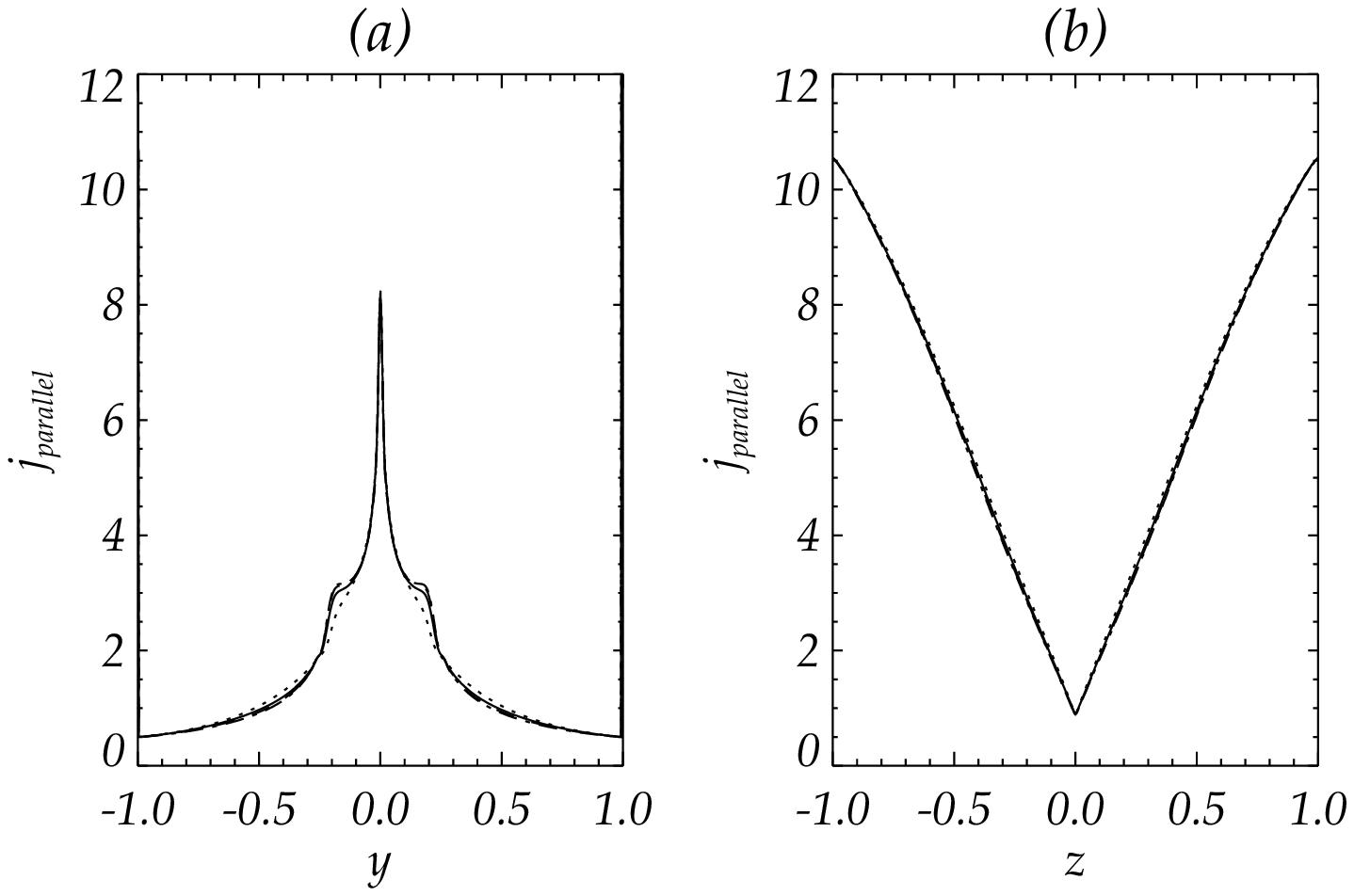}
  \caption{Plot of the parallel (to the magnetic field) component of the electric current density (a) in a horizontal cut $x=0$ plane, at height $z=0.7$, and (b) along the spine axis ($z$-axis), for four different experiments with $p_0=0.1$ (dotted), $p_0=0.5$ (dashed), $p_0=1$ (solid) and $p_0=1.5$ (dash-dotted).}
  \label{fig:varyp}
  \vspace{0.3cm}
  \includegraphics[scale=0.30]{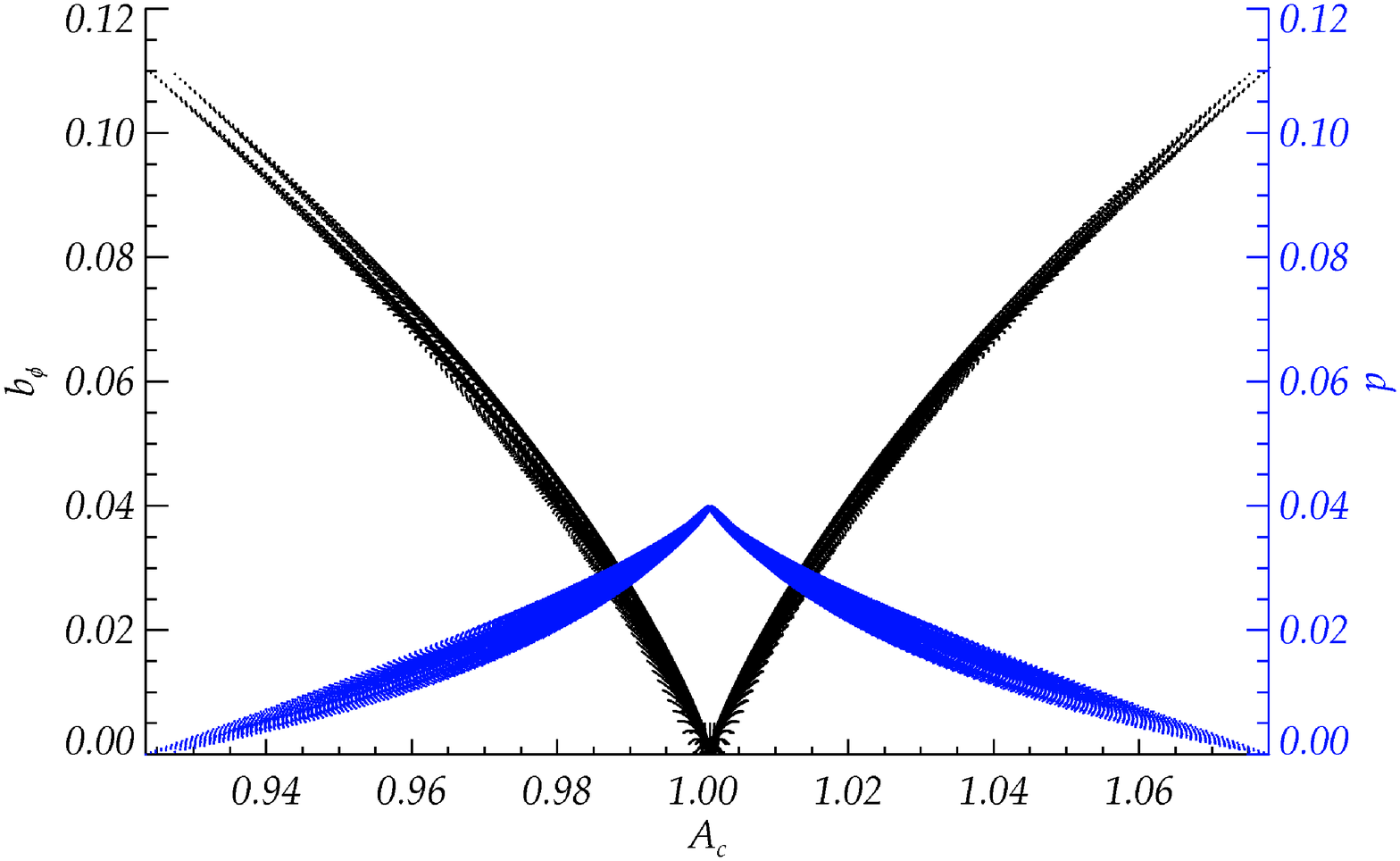}
  \caption{Numerical dependence of poloidal field (black) and plasma pressure (blue) with the function $A_c$. All points of a region about the null that goes from -0.5 to 0.5 both in $r$ and $z$ are plotted here.}
  \label{fig:sp_ff1}
  \vspace{0.3cm}
\end{figure}

Finally, a comparison has been made of results with different background plasma pressures, with $p_0=0.1,0.5,1.0,1.5$. In Fig. \ref{fig:varyp} we show the parallel electric current density through a horizontal cut of Fig. \ref{fig:sp_current}a, at $z=0.7$, and in a vertical cut along the spine ($z$-axis). The plots show the results for the four different initial plasma pressures, but since they all overlap, only one line is visible, with the exception of the smallest pressure case ($p=0.1$), which shows a slightly smoother profile. Hence, changes in the plasma pressure do not appear to lead to significant changes in the formation of the current concentrations.


\subsection{The 2.5D Approximation} \label{sec:2halfD}

As has been discussed above, close to the null point, the field lines expands radially and the system can be assumed to have cylindrical symmetry about the axis of the spine. This symmetry breaks slightly when we approach the edges of the box, due to the squared line-tied boundaries. Under the assumption of cylindrical symmetry, all the quantities depend only on two coordinates, namely, the height, $z$, and the radius, $r$, and the equilibrium can be described as a 2.5 dimensional state (that is, there are three spatial coordinates, but all the quantities only depend on two of them).

Using the solenoidal constraint (i.e. $\boldnabla\cdot{\bf B}=0$), the magnetic field for a two-dimensional cylindrical system must satisfy,
\begin{equation}
\frac{1}{r}\frac{\partial}{\partial r}(rB_r)+\frac{\partial B_z}{\partial z}=0\;.
\end{equation}
Also, if $\partial/\partial\phi=0$, the magnetic field, ${\bf B}=(B_r,B_{\phi},B_z)$, may be rewritten as
\begin{equation}
{\bf B}(r,z)=\frac{1}{r}\boldnabla A_c(r,z)\times{\bf e}_{\phi}+B_{\phi}(r,z){\bf e}_{\phi}\;,
\end{equation}
where $A_c(r,z)$ is a function analogous to the flux function in 2D, for systems with cylindrical symmetry, which satisfies
\begin{equation}
(B_r, B_{\phi}, B_z)=(-\frac{1}{r}\frac{\partial A_c}{\partial z},\; \frac{b_{\phi}(A_c)}{r},\; \frac{1}{r}\frac{\partial A_c}{\partial r})\;,
\end{equation}
where $b_{\phi}(A_c)$ is an unknown function that depends uniquely on $A_c$.

As in the 2D case, in a magnetohydrostatic (MHS) equilibrium that satisfies these conditions, the plasma pressure must be a unique function of $A_c$, as well as $b_{\phi}$. Therefore, a Grad-Shafranov equation can be derived, which defines the force balance between the plasma and magnetic forces in final equilibrium, as
\begin{equation}
\frac{1}{\mu_0}\left[\frac{\partial^2 A_c}{\partial r^2}-\frac{1}{r}\frac{\partial A_c}{\partial r}+\frac{\partial^2 A_c}{\partial r^2}\right]+b_{\phi}\frac{{\rm d}b_{\phi}}{{\rm d}A_c}+\frac{{\rm d}p}{{\rm d}A_c}=0\;.
\end{equation}

Fig. \ref{fig:sp_ff1} shows the numerical functionality of the poloidal field function $b_{\phi}(A_c)$ and the plasma pressure $p(A_c)$ of the final equilibrium state, for all the points within a region about the null (to avoid non-cylindrical asymmetries) that goes from -0.5 to 0.5 both in $r$ and $z$. Both quantities show a good dependence in this region, and the term of the poloidal field, $b_{\phi}$. In particular, it shows that, in the region about the null, where the boundaries do not have a strong effect, a non-force-free description is found, where the plasma pressure plays an important role.

The assumption of axial symmetry about the spine is non-generic. It is very likely that some sort of asymmetry in the field would arise in a realistic situation. In the next section, we consider the current structure generation due to the collapse of non-axi-symmetric spiral nulls, where the field lines have a preferred direction in the fan plane, and therefore, the 2.5D assumption is not valid even close to the null point.


\section{Magnetic field asymmetries} \label{sec5}

\begin{figure*}[t]
  \centering
  \includegraphics[scale=0.65]{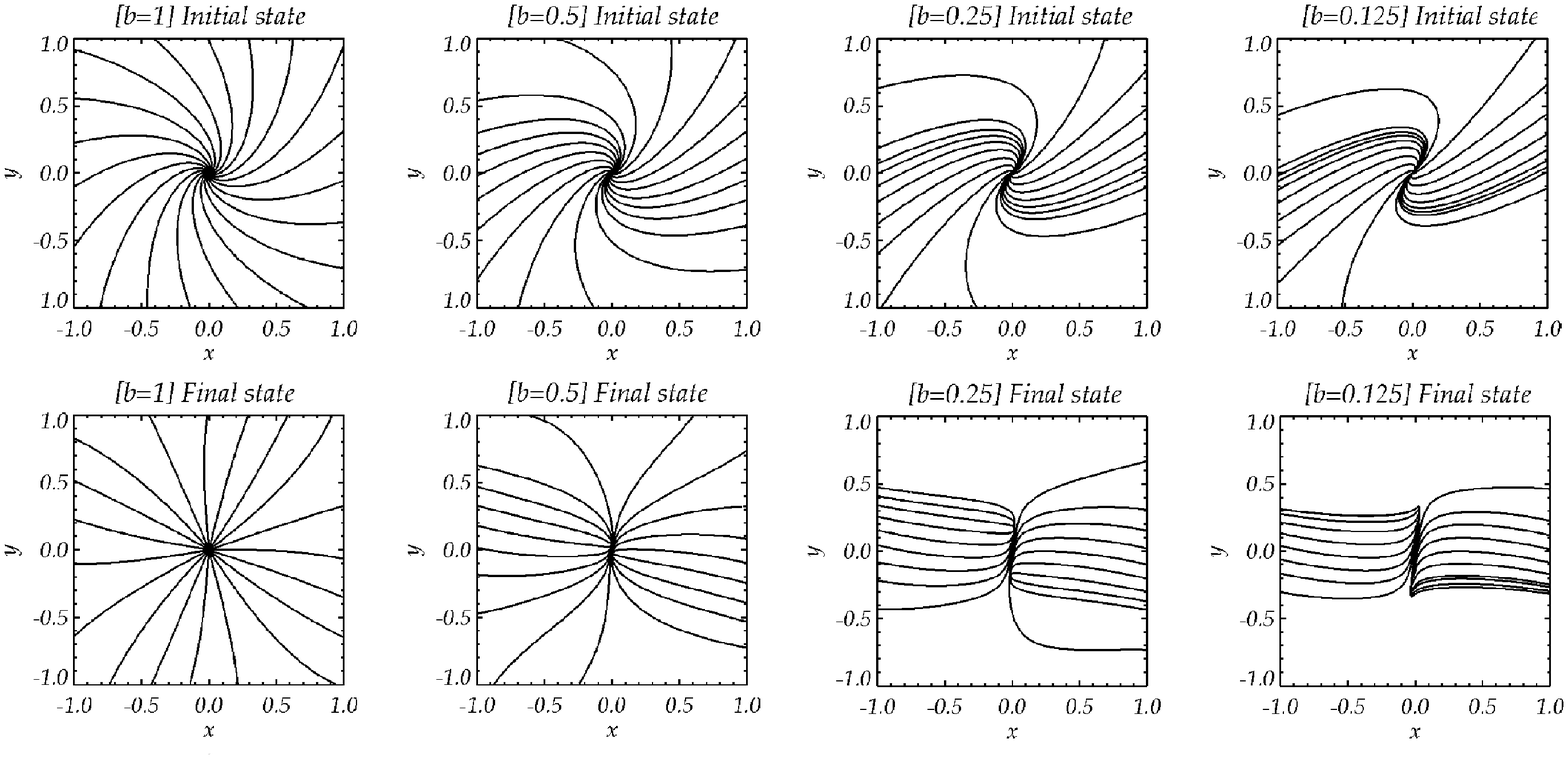}
  \caption{Magnetic field lines lying on the fan surface for the initial (top row) and final equilibrium (bottom row) state, for four different experiments, with, from left to right, $b=1, 1/2, 1/4, 1/8$, and $j_{sp}=1$.}
  \label{fig:fanfieldlines}
  \vspace{0.3cm}
\end{figure*}

\begin{figure*}[t]
  \centering
  \includegraphics[scale=0.65]{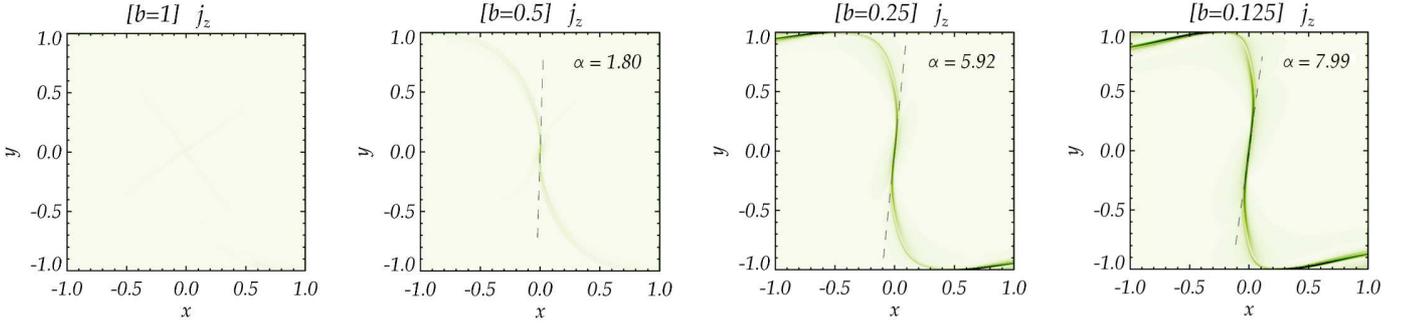}
  \caption{Current density on the fan plane for the final state, for the same four experiments of Fig. \ref{fig:fanfieldlines}. The central current layer extends along the direction of one of the eigenvectors (the minor axis) associated with the linear analysis of the initial magnetic field about (dashed lines), where $\alpha$ is the angle (in degrees) this eigenvector makes with the $y$-axis. The maximum current density value in these plots (dark green) is $j_z=20$, with $j_z=0$ as the minimum (white).}
  \label{fig:fancurr}
  \vspace{0.3cm}
\end{figure*}

By removing the constraint that $b=1$ in Eq. (\ref{M_general}), we can introduce axial asymmetries to the initial state. Hence, from Eq. (\ref{M_general}) our initial magnetic field is of the form
\begin{equation}
(B_x, B_y, B_z)=(x-\frac{j_{sp}}{2}y, \frac{j_{sp}}{2}x+by\;, -(b+1)z)\;. \label{genb}
\end{equation}

Depending on the value of $b$ relative to the magnitude of the initial current density, $j_{sp}$, the magnetic field lines can take different shapes. According to \citet{Parnell96}, we can define a threshold for the current density, which for our field is $j_{thresh}=|b-1|$, above which the fan field lines are spiral, below which they are said to be of skewed improper form. We consider here the case where $|j_{sp}|>|b-1|$. The fan field lines form a spiral null, as in the $b=1$ axi-symmetric case. If $b\ne 1$, this spiral will bend towards a preferred direction (in our case, the preferred direction is the $x$-axis if $0<b<1$, or the $y$-axis if $b>1$).

Previous studies on the effect of magnetic field asymmetries in 3D spiral nulls have been considered by \citet{Pontin11}. They focus on the reconnection rate for both torsional spine and torsional fan reconnection, but do not look at the formation of equilibrium current structures. In the particular torsional spine case, they start with a potential configuration with different values of $b$, and then introduce a localised magnetic field perturbation in the form of a ring of magnetic flux. They find that the peak current and the reconnection rate do not depend strongly on the degree of asymmetry (given by the value of $b$). This dependence is shown to be more important for torsional fan reconnection case (which is not considered here).

Equation (\ref{genb}) gives a generic form for the magnetic field around a 3D magnetic null point with a homogeneous current parallel to the spine line \citep{Parnell96}. The case of $b=1$ makes $j_{thresh}=0$, and hence, corresponds to the axi-symmetric case studied in Sec. \ref{sec4}, where the two eigenvectors lying on the fan plane are perpendicular to each other, and their eigenvalues are complex conjugates. This is not the case for the asymmetric fields, where the eigenvectors are no longer perpendicular, although their eigenvalues are still complex conjugates. Note, from Eq. (\ref{M_general}), that the effect of increasing or decreasing $b$ by the same amount (i.e. multiplying or dividing by the same value) is analogous. The only difference in doing so is the preferred axis that the magnetic field lines will follow, being either $x$ or $y$ (i.e. a rotation of $90^{\circ}$ around the spine axis). Therefore, without lost of generality, we have chosen only values of $b$ smaller than $1$. We run relaxation experiments for various values of $b<1$, with a fixed current parallel to the spine, $j_{sp}=1$. As before, during the relaxation the energy is checked to ensure that it is conserved to within numerical error and we note that a similar amount of viscous dissipation occurs, resulting in a rise in internal energy of order 1\%.

In Fig. \ref{fig:fanfieldlines} we show the fan plane field lines in the initial non-equilibrium and the final equilibrium state for four different experiments with four different values of $b$, namely, $b=1, 1/2, 1/4, 1/8$, and with $j_{sp}=1$ for all of them. As the parameter $b$ is systematically decreased, the fan field lines show a more pronounced bend towards the $x$-axis, both in the initial and in the final state. Initially, the asymmetry appears as a skewed spiral, and after the dynamical relaxation, the fan field lines show a more complex geometry.

\begin{figure}[t]
  \centering
  \includegraphics[scale=0.30]{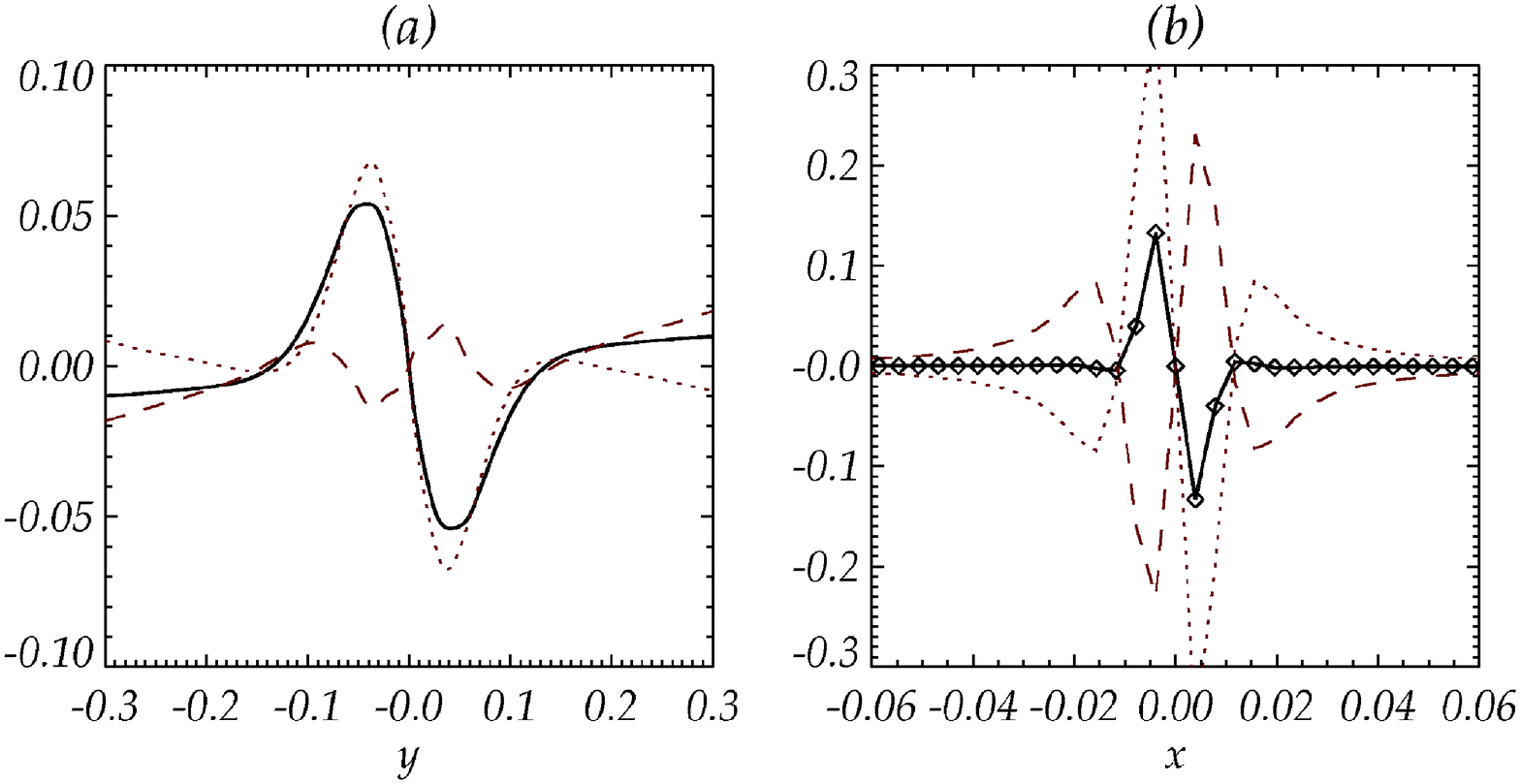}
  \caption{Forces in the final state, in the fan plane, for the case $b=0.25$, showing (a) the $y$-component along the $y$-axis (along the current accumulation) and (b) the $x$-component along the $x$-axis, of the Lorentz forces (dotted), the pressure forces (dashed) and the total forces (solid).}
  \label{fig:nosy_forces}
  \vspace{0.3cm}

  \centering
  \includegraphics[scale=0.75]{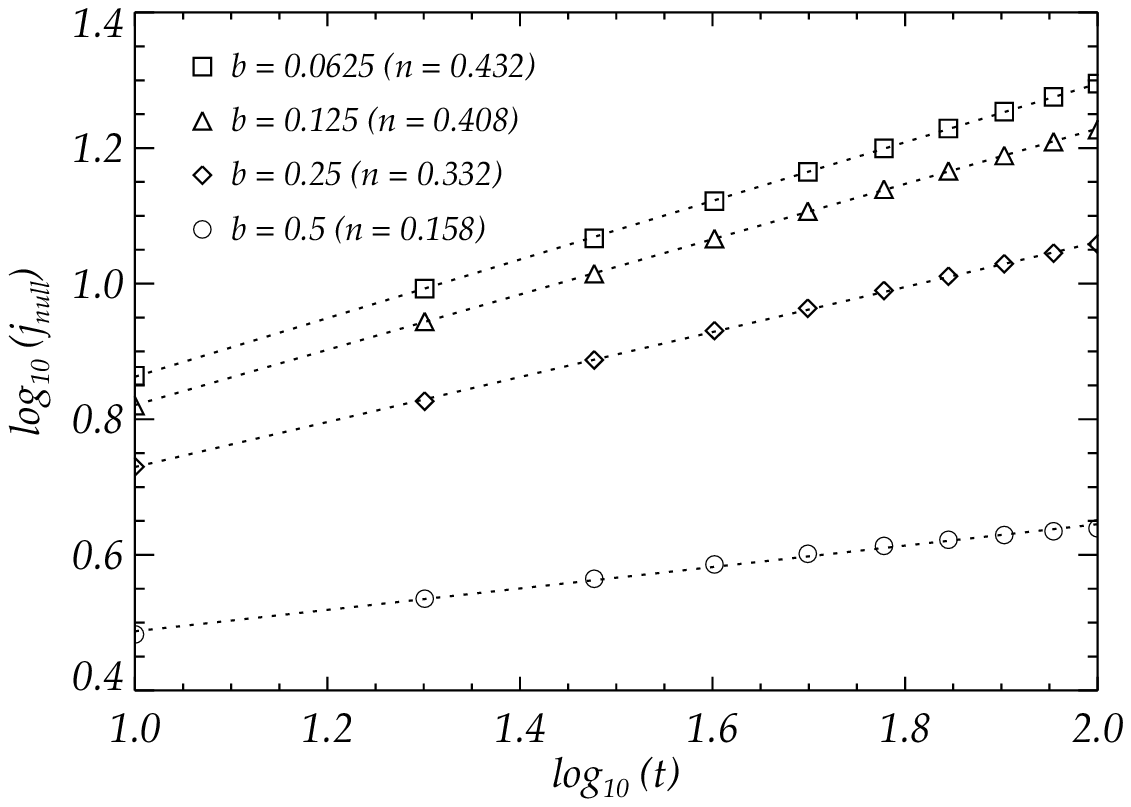}
  \caption{Time evolution, with logarithmic axes, of the null point current current density for four asymmetric experiments, with $b=1/2, 1/4, 1/8$ and $1/16$. Dotted lines show fits to $j_{null}=mt^n$.}
  \label{fig:nullj}
  \vspace{0.3cm}
\end{figure}

To evaluate the deviations from the potential field in the final state, we look at the current accumulations at the fan plane. The only non-zero component of the current density for all cases is checked to be $j_z$. This implies that the current density at the null is parallel to the spine, and hence, the spine and the fan plane remain perpendicular to each other. In Fig. \ref{fig:fancurr} we show the $z$-component of the current density for the same four experiments as in Fig. \ref{fig:fanfieldlines}. The final state is a non-force free equilibrium, as in the axi-symmetric case, but here, a new feature is formed along one of the eigenvectors of the fan plane (known as the minor fan axis). This is formed because the initial forces in the fan plane are not axi-symmetric and are weaker along the minor axis. Note, that in the symmetric case, the only non-zero initial force is the magnetic tension force, which acts axi-symmetrically, while here, there is also a magnetic pressure force. The overall Lorentz force in the fan plane is such that it prevents the field from evolving towards a potential configuration. Instead, it accumulates the current in a current layer through the null itself.

Again, by considering the forces, we find that in the final state, the system is in a non-force-free equilibrium everywhere except very close to the null, where tiny residual forces remain on the fan plane. These are shown in Fig. \ref{fig:nosy_forces}, for two cuts along the $y$ and $x$ axes. The Lorentz forces are directed towards the null. The pressure forces act against the Lorentz forces but are not able to balance them. This results in a non-zero net total force which causes the gradual accumulation of current at the null point and is consistent with the formation of a singularity whose nature is discussed shortly. These results are in agreement with \citet{Parnell97}, which found that the linear field about a non-potential null cannot be balanced by a pressure gradient. Hence, since the field in the vicinity of the null is not potential, an equilibrium cannot be achieved there. Moreover, the magnetic field in the fan plane is weaker along the minor axis of the null, thus, as already mention, too are the Lorentz forces, and therefore that is the axis along which the current density is able to extend.

The larger the initial field asymmetry, the larger the amplitude of the null current layer. Note, this current layer is a feature of the asymmetry, and does not appear in the axi-symmetric case. Also, it is not related to a fan-spine collapse, as the current density about the null remains parallel to the spine. It is noticeable how, for the largest degree of asymmetry, the null current density is the maximum current of the domain.

The creation of a pronounced layer of current density perpendicular to the spine, at the location of the magnetic null point and along the minor fan axis, and the associated forces plotted in Fig. \ref{fig:nosy_forces}, suggest the formation of a singularity \citep{Craig05,Pontin05b,Fuentes11}. Figure \ref{fig:nullj} shows the time evolution of the current density at the null point for four different asymmetric experiments. Each of these evolutions is modelled with a function of the form
\begin{equation}
j_{null}=mt^n\;,
\end{equation}
where both $m$ and $n$ are positive and both depend on the degree of asymmetry, defined by $b$. The null current density seems to have entered an asymptotic regime similar to the one studied by \citet{Fuentes11} for 2D X-points. The results shown in Fig. \ref{fig:nullj} suggest the formation of an infinite-time singularity at the null point, associated with a creation of the perpendicular current density caused simply by an initial asymmetry of the magnetic configuration. Note, that the appearance of such a singularity is strictly linked to the asymmetric nature of the initial state. This singularity does not exist if the field is axi-symmetric, because in that case, the field can become potential locally about the null point.

Current singularities at three-dimensional magnetic null points have been found after the fan-spine collapse of a tilted null \citep{Pontin05b}. The key difference between these singularities and the ones described in this paper is the direction of the current density vector. In the fan-spine collapse, the current is directed along the tilt axis of the fan, for example, the $x$-axis \citep{Pontin05a,Pontin07a,Pontin07b,Pontin07c,Masson09,Masson12}. On the contrary, our current density is directed along the spine line, i.e. the $z$-axis. Therefore, our singularity is not due to a collapse of the fan and the spine towards one another, but is a completely new feature that has never been observed in the past. This study is the first signature of a current density singularity at a purely spiral null, and further studies are needed to show if it is associated with a new regime of 3D null reconnection.


\section{Summary and conclusions} \label{sec6}

The three-dimensional relaxation of spiral magnetic null points with initial spine-aligned current has been investigated under non-resistive conditions, resulting, in all cases, in a non-force-free equilibrium, where all the velocities have been damped out by the viscous forces. In the final states, forces of the plasma pressure and magnetic field, which are of about the same order as the initial non-equilibrium Lorentz forces, balance each other in a genuine non-force-free state.

In the axi-symmetric case, the initial field is a spiral null with a uniform twist of the field lines everywhere. During the relaxation, the field evolves by concentrating the initial constant current density around the spine, above and below the fan, and hence, the twist of the field lines also resides there. The fan field lines show a radial configuration, with small departures from the radial field caused by the squared line-tied boundaries. In the final equilibrium, we decompose the electric current density vector into two components, parallel and perpendicular to the magnetic field, and we find that the parallel component is about ten times higher than the perpendicular component. This means that the field is very close to a force-free field, although cannot exactly be described as such.

The parallel current, which is the dominant component, is maximum along the spine line, and draws an hourglass shape with two cones that meet at the location of the null. The accumulations of perpendicular current (regions of non-force-free-ness) occur at the edges of the two cones about the spines, above and below the fan. In the axi-symmetric case, the field is shown to be potential only for the small region about the null, where $r<0.1$. Systematic changes of the initial plasma pressure have been studied, finding the same amplitudes for the current accumulations in all cases. This indicates that changes in the plasma pressure may not play an important role in the evolution of axi-symmetric spiral nulls.

Also, the magnetic field in the axi-symmetric case shows cylindrical symmetry about the spine for $r<0.5$. A Grad-Shafranov equation can be used in this case to describe the field analytically, demonstrating the importance of the plasma pressure force to balance the Lorentz force.

Overall, the final equilibrium for the initially axi-symmetric null point shows three distinct regions, i) a potential envelope surrounding the null, at $r<0.1$, ii) an axi-symmetric region where the system is in non-force-free equilibrium ($0.1<r<0.5$), and iii) the region close to the boundaries, where the cylindrical symmetry is broken by the squared line-tied boundaries, but where the system is still able to achieve a non-force-free equilibrium.

By breaking the initial axial symmetry of the field lines, we obtain new features that were not present in the axi-symmetric case. The fan field lines are not radial any more. Instead, they bend towards a preferred direction, and they build a pronounced perpendicular current accumulation at the null, elongated along the direction of the minor axis, about which the magnetic field strength is smaller. In these cases, locally about the null, the plasma pressure forces are not able to balance the Lorentz forces, which are directed towards the null forming an infinite-time singularity. This is not present in the axi-symmetric case, since it can reach a potential equilibrium very close to the null, save exactly at the null itself. The current at the null is directed along the spine axis and the spine and fan remain perpendicular, therefore, it is not caused by a fan-spine collapse, like the commonly known current singularities previously found at three dimensional null points \citep{Pontin05b}. The kind of singularity observed in this paper for the asymmetric spiral nulls has, to our knowledge, never been observed in the past. It is proposed that this new singularity, which arises naturally from the non-symmetric character of our initial state, may be linked to a new regime of magnetic reconnection at 3D magnetic nulls.

The MHD evolution of tilted nulls with initial fan-aligned current, the effects of plasma pressure, and the formation of a current accumulations such as the one studied by \citet{Pontin05b}, will be studied in a follow-up paper from the series of {\it dynamical relaxation of coronal magnetic fields}.


\section*{Acknowledgements}

The authors would like to thank the referee for many useful comments which helped to understand the details of the experiments carried out in this paper. This work has been partly supported by SOLAIRE European training network. Computations were carried out on the UKMHD consortium cluster funded by STFC and SRIF. JFF is funded from the St Andrews Rolling Grant (ST/H001964/1).


\bibliographystyle{aa}
\bibliography{jfuentes3}

\end{document}